\documentstyle[11pt]{article} 
\begin{document} 
\centerline{\bf Quantum cryptography based on photon ``frequency'' states:}
\centerline{\bf example of a possible realization} 
\centerline{\bf S. N. Molotkov}
\centerline{\bf Institute of Solid-State Physics, Russian Academy of Sciences,}
\centerline{\bf 142432 Chernogolovka, Moscow District} 
E-mail:molotkov@issp.ac.ru 

\begin{abstract}
A quantum cryptosystem is proposed using single-photon states with
different frequency spectra as information carriers.  A possible
experimental implementation of the cryptosystem is discussed.
\end{abstract}

\section*{1.~~Introduction}

The idea of quantum cryptography was first proposed in
Ref.\cite{1}, initially inaccessible.  In its current form the
protocol for propagation of a key (a secret random sequence of zeros and
ones) was proposed in Ref.~\cite{2}.  A new qualitative jump in the
understanding of secrecy in quantum cryptography arose after
Ref.~\cite{3}, in which a protocol was proposed for exchange using
nonorthogonal states, and Ref.~\cite{4} where a protocol was
described using the Einstein--Podolsky--Rosen effect.

Various versions of cryptosystems using nonorthogonal states and the
Einstein--Podolsky--Rosen effect were proposed
subsequently.\cite{5,6,7,8,9,10,11,12,13,14,15,16,17,18,19,20,21,22}  
Experimental prototypes of quantum cryptosystems have been implemented
using nonorthogonal polarization states of photons for
coding\cite{5,11,12} and also using the principle of phase coding based on
a fiber-optic Mach--Zehnder time-division interferometer 
\cite{8,9,10,19} (an improved version of the system from
Ref.~\cite{10} was implemented in Ref.~\cite{19}).  The
longest communication channel achieved under laboratory conditions is
30\,km (Ref.~\cite{10}).  The operation of a prototype quantum
cryptosystem has been demonstrated under natural conditions using a 23\,km
long optical cable below Lake Geneva\cite{18} and also between various
buildings at the Los Alamos National Laboratory.\cite{13}

Most of these systems use the interference principle which roughly
involves ``splitting'' a photon at the transmitting end of the line and
``collimating'' it at the receiving end.  Here we propose a quantum
cryptosystem using different photon frequency states which utilizes the
``internal interference'' of the different photon frequency components.
Such a system may well prove more stable in operation than direct
interference systems although this can only be confirmed by means of an
experimental implementation.

The secrecy of the key in quantum cryptography is based on two facts:  
1) the impossibility of coding (cloning) a previously unknown quantum
state \cite{23} and 
2) the impossibility of extracting information on quantum states without
perturbing them if they belong to a nonorthogonal basis.\cite{3} 
Formally, any pair of nonorthogonal states corresponding to logic 0 and 1
can be used as information carriers.  The detection procedure
(quantum-mechanical measurement) at the receiving end should be set up
such that any attempts at intervention  within the communication channel,
i.e., changes in states, can be identified from the results of the
measurements.  If the pair of nonorthogonal states $|\psi_0\rangle$ and
$|\psi_1\rangle$ are used as carriers, the formal measurements are given
by the projectors\cite{3} 
\[
\overline{E}_0=1-|\psi_0\rangle\langle\psi_0|,\quad 
\overline{E}_1=1-|\psi_1\rangle\langle\psi_1|,
\]
whose action reduces to the projection of states orthogonal to the vectors
$|\psi_0\rangle$ and $|\psi_1\rangle$, respectively, on the
subspaces.\cite{3}  The result of the action of the projectors is treated
as a statement and the probability of the measurement results is given by
the expressions:
\begin{eqnarray}
\mbox{Pr}&=&\mbox{Tr}\{\hat{\rho}_0\overline{E}_0\}=
\mbox{Tr}\{\hat{\rho}_1\overline{E}_1\}\equiv 0,
\nonumber \\
\mbox{Pr}&=&\mbox{Tr}\{\hat{\rho}_0\overline{E}_1\}=
\mbox{Tr}\{\hat{\rho}_1\overline{E}_0\}=
1-|\langle\psi_0|\psi_1\rangle|^2\neq 0.
\end{eqnarray}
Measurements using $\overline{E}_0$ and $\overline{E}_1$ in an ideal
communication channel (without noise) can detect any attempts at
eavesdropping, i.e., changes in states.  The first nonzero outcome of a
control measurement definitely indicates the presence of an eavesdropper.
When it is known that a signal, say $|\psi_0\rangle$, was sent,
measurements were made using $\overline{E}_0$, and a nonzero result
obtained, the nonzero result is considered to be a statement that the
state $\overline{E}_0$ has a nonzero component in the corresponding
orthogonal complement of the Hilbert space (i.e., that the state
$|\psi_0\rangle$ was changed).

Photon states are used as information carriers for real fiber-optic
communication channels.  Formally any pair of nonorthogonal photon states
can be used (not necessarily even single-photon states).  However, it is
unclear how the measurement procedure corresponding to the projectors
$\overline{E}_{0,1}$ for a given pair of states can be implemented
experimentally.  It would be easiest to use a pair of nonorthogonal
polarizations but an optical fiber does not ``support'' polarization (see
details given in Refs.~\cite{11} and \cite{12}).  The
prototypes of quantum cryptosystems mentioned use the phase coding
principle based on time-division interferometer.\cite{8,9,10}
After the interferometer has been operating for a few minutes, the system
requires additional alignment.\cite{10}

\section*{2.~~Cryptography using photon ``frequency'' states}

A quantum cryptographic system using the Einstein--Podolsky--Rosen effect
for a biphoton field was proposed in an earlier study\cite{20} and this
idea is developed here.  Subsequently, the use of frequency states which
do not use interference over large distances is suggested.
Let us first analyze a formal system and then discuss an experimental
realization.  Three single-photon states are used as carriers:  two
information states corresponding to logic zero and one and one control
state.  The information states are mutually orthogonal.  The control state
is pairwise nonorthogonal to the information states.  The use of only two
information states is inadequate for secrecy because they can be reliably
distinguished as a result of their orthogonality.
The information states comprise pure stationary states with the density
matrices
\begin{equation}
\hat{\rho}_0=|e_0\rangle\langle e_0|, \quad
\hat{\rho}_1=|e_1\rangle\langle e_1|, \quad 
\langle e_1|e_0\rangle=0, 
\end{equation}
where $|e_0\rangle$ and $|e_1\rangle$ are certain basis states assigned to
the energies $\omega_{0}$ and $\omega_{1}$, respectively.  The control
state is nonstationary and contains both basis components 
$|e_0\rangle$ and $|e_1\rangle$:
\begin{eqnarray}
|\psi_c(t_0)\rangle&=&
\mbox{e}^{-i\omega_0t_0}f_0|e_0\rangle+
\mbox{e}^{-i\omega_1t_0}f_1|e_1\rangle,\nonumber \\
\hat{\rho}_c(t_0)&=&|\psi_c(t_0)\rangle\langle \psi_c(t_0)|, 
\end{eqnarray}
and the normalization condition
\begin{displaymath}
|f_0|^2+|f_1|^2=1.
\end{displaymath}
The time $t_{0}$ describes the beginning of the time measurement --- the
state preparation time (see below).  The density matrix at times $t >t_{0}$ 
is obtained by substituting into the argument
$\hat{\rho}_{c}(t)=|\psi_c(t-t_0)\rangle\langle\psi_c(t-t_0)|$.  The
introduction of two orthogonal information states reduces the number of
``idle'' outcomes because of their distinguishability if no eavesdropping
is detected in the exchange process.
This scheme uses two types of measurements.  The measurements of the
frequency spectrum are described by an orthogonal resolution of identity in
space spanned on states $|e_{0}\rangle$, $|e_{1}\rangle$:
\begin{equation}
E_0+E_1=I,
\quad E_{0}=|e_{0}\rangle\langle e_{0}|,
\quad E_{1}=|e_{1}\rangle\langle e_{1}|,
\end{equation}
where $I$ is the unit operator.  The second family of measurements
involves measuring the time and is given by a nonorthogonal resolution of
identity (see Ref.~\cite{24}, for example), which in our case has the
form
\begin{eqnarray}
\int\limits_{0}^{T}E(dt)=I, \quad
T&=&\frac{2\pi}{|\omega_1-\omega_0|},
\nonumber \\
E(dt)&=&
\left(
\mbox{e}^{-i\omega_0t}|e_0\rangle+\mbox{e}^{-i\omega_1t}|e_1\rangle
\right)
\left(
\langle e_0|\mbox{e}^{i\omega_0t}+\langle e_1|\mbox{e}^{i\omega_1t}
\right)
\frac{dt}{ T}.
\end{eqnarray}

In accordance with the general philosophy of quantum-mechanical
measurements, the measurements were made at a certain time.\cite{24,25,26}
The probability of the outcome of the measurements using the projectors
$E_{0}$ and $E_{1}$ does not depend on time and is given by
\begin{eqnarray}
\mbox{Pr}&=&\mbox{Tr}\{\hat{\rho}_{0}E_0\}=1,\quad
\mbox{Pr}=\mbox{Tr}\{\hat{\rho}_{1}E_1\}=1,\quad
\mbox{Pr}=\mbox{Tr}\{\hat{\rho}_{0,1}E_{1,0}\}\equiv 0,
\nonumber \\
\mbox{Pr}&=&\mbox{Tr}\{\hat{\rho}_{c}(t)E_0\}=|f_0|^2,\quad
\mbox{Pr}=\mbox{Tr}\{\hat{\rho}_{c}(t)E_1\}=|f_1|^2.
\end{eqnarray}
Measurements of the time give the probability distribution of the outcomes
in the range $(t,t+dt)$:
\begin{equation}
\mbox{Pr}(dt)=\mbox{Tr}\{\hat{\rho}_{0,1}E(dt)\}=1\cdot 
\frac{ dt}{ T},
\end{equation}
\vspace{-4mm}
\begin{eqnarray}
\mbox{Pr}(dt)&=&\mbox{Tr}\{\hat{\rho}_c(t)E(dt)\}=
\left|f_0
\exp \left[-i\omega_0(t-t_0)\right]+f_1
\exp \left[-i\omega_1(t-t_0)\right]
\right|^2
\left( \frac{ dt}{ T}\right)
=
                \nonumber\\
&=&\left\{
1+2\mbox{Re}
\left[
f_0f_{1}^{*}\exp \left[-i(\omega_0-\omega_1)(t-t_0)\right]
\right]
\right\}
\left( \frac{dt}{ T} \right).
\end{eqnarray}
For the control state the probability is an oscillating function with the
period $T=2\pi/|\omega_1-\omega_0|$.  This set of measurements can
completely reconstruct information on the states --- no other density
matrices can reproduce the statistics of the measurements so that any
attempts at eavesdropping can be detected (for further details see
Ref.~\cite{22}).

The key generation protocol is as follows.  We assume that all the
parameters of the states are known to everybody, including any potential
eavesdropper.  User $A$ (henceforth ``Alice'') randomly sends into the
communication channel states $\hat{\rho}_c$, $\hat{\rho}_0$, or
$\hat{\rho}_1$.  User $B$ (henceforth ``Bob'') randomly and independently
of Alice selects measurement type $E_0$, $E_1$, or $E(dt)$. After making 
a 
series of measurements, Alice transmits through the open channel
(accessible to all including an eavesdropper ---``Eve'') the numbers of
some measurements when $\hat{\rho}_0$ and $\hat{\rho}_1$ were sent and all
the numbers when control state $\hat{\rho}_c$ was sent.  Bob sorts the
measurements into three groups according to when $\hat{\rho}_c$,
$\hat{\rho}_0$, or $\hat{\rho}_1$ were transmitted. In each of these three
groups, three subgroups are identified according to measurement procedures
$E_{0}$, $E_{1}$, or $E(dt)$.  For instance, for those messages when Alice
transmitted state $\hat{\rho}_c$, the relative fraction of the measurement
outcomes when the projectors $E_{0}$ and $E_{1}$ were used should be
$|f_0|^2/|f_1|^2$ regardless of the measurement time.  For the $E(dt)$
measurements the probability of the measurement results at various times
should converge toward the probability distribution (8).  The convergence
of the distribution function for a finite sample should by checked by
using some statistical criterion such as the Kolmogorov criterion\cite{27}
(see also Ref.~\cite{22}).  The convergence is checked similarly for
the measurements when states $\hat{\rho}_0$ or $\hat{\rho}_1$ were sent.

For example, for state $\hat{\rho}_0$ the measurements using $E_{0}$ should 
give the same outcome in all attempts, which does not depend on the
measurement time.  For the $E_{1}$ measurements in all attempts the
outcome should be zero regardless of the measurement time.  For the
$E(dt)$ measurements the probability of the outcome is only determined by
the duration of the time interval $dt$ and does not depend on the time
$t$.

The secrecy of the protocol is guaranteed by the nonorthogonality of the
information states to the control state and by the fact that a set of
measurements is information-complete so that any attempts at
eavesdropping, i.e., changes in states, can be detected.  In other words,
no other density matrices can reproduce the statistics of the measurements
at the receiving end (for further details see Ref.~\cite{22}).  

After having established that no eavesdropping is taking place, Alice
transmits the numbers of those measurements when the control state was
sent. 
All the idle measurements when the detector was not triggered are
discarded.  Then, for the remaining numbers Bob only transmits the numbers
of those measurements in which he used $E_{0}$ or $E_{1}$ but does not
communicate which measurement, $E_{0}$ or $E_{1}$, was used in each
specific attempt (this information is now known only to Alice and Bob).

These remaining measurements give the secret key (an identical random
sequence of zeros and ones for Alice and Bob). 

We shall illustrate why an eavesdropper will inevitably introduce errors.
In order to obtain information on the key,  Eve must distinguish states
$\hat{\rho}_0$ and $\hat{\rho}_1$.  To do this, she must make measurements
with a narrow-band detector (measurements of $E_{0}$ or $E_{1}$).  If
there were no control state $\hat{\rho}_c$ containing both spectral
components with frequencies $\omega_{0}$ and $\omega_{1}$, as a result of
the mutual orthogonality of the information states, it would be possible
to determine uniquely which state is present in the line.  However, their
nonorthogonality to the control state will inevitably lead to errors since
there will always be measurements with an undetermined result.  For 
instance, if $\hat{\rho}_c$ is present in the line and Eve measured 
$E_{0}$ and obtained a nonzero result, it is impossible to uniquely
determine which state, $\hat{\rho}_c$ or $\hat{\rho}_0$, gave this result
Resending $\hat{\rho}_0$ instead of the true state $\hat{\rho}_c$ leads to
a change in Bob's measurement statistics.  It is also impossible to
discern in one measurement that both spectral components with energies
$\omega_{0}$ and $\omega_{1}$ are present simultaneously in a state
because of the orthogonality of the components since this requires
measurements by means of $E_0 E_1$.  This projector can be considered as
confirmation that the property $E_{0}$ ($\omega_{0}$ present) and $E_{1}$
($\omega_{1}$ present) are found simultaneously.  However because of the
orthogonality ($E_0\bigcap E_1=\oslash$) the action of $E_0 E_1$ on any
density matrix has a result identically equal to zero.  Also no unique
information can be obtained on the simultaneous presence of spectral
components using more general (non-von Neumann)  measurements,  which is
guaranteed by the theorem in Ref.~\cite{3}.

\section*{3. Possible implementation of a cryptosystem}

We shall now discuss a possible experimental implementation in which the
carriers are three single-photon states of the type
\begin{eqnarray}
|1_{\omega_0}\rangle&=&
a^{+}_{{\small\mbox{\boldmath $\epsilon$}},\omega_0}|0\rangle,\quad
|1_{\omega_1}\rangle=
a^{+}_{{\small\mbox{\boldmath $\epsilon$}},\omega_1}|0\rangle, 
\nonumber \\
|1_{c}\rangle&=&
f_0\mbox{e}^{-i\omega_0 t_0 }a^{+}_{{\small\mbox{\boldmath $\epsilon$}},
\omega_0}|0\rangle+
f_1\mbox{e}^{-i\omega_1 t_0}   
a^{+}_{{\small\mbox{\boldmath $\epsilon$}},\omega_1}|0\rangle
\end{eqnarray}
with the corresponding density matrices
\begin{displaymath}
\hat{\rho}_{0,1}=|1_{\omega_{0,1} }\rangle\langle 1_{\omega_{0,1}}|,\quad\hat{\rho}_{c}=|1_{c}\rangle\langle 1_{c}|,
\end{displaymath}
where $a^{+}_{{\small\mbox{\boldmath $\epsilon$}},\omega_i}$ is the
creation operator of a Fock monochromatic state with the frequency
$\omega_i$ ($i,1$) and polarization $\mbox{\boldmath $\epsilon$}$,  
and $|0\rangle$
is the vacuum state.  Quite clearly, a strictly monochromatic state is 
an idealization.  However, there are no fundamental constraints on the
formation of states arbitrarily close to monochromatic.

The measurement procedures described above may be implemented by using a
fast (fairly wide-band) photodetector operated in a waiting regime, and
two narrow-band filters at frequencies $\omega_{0}$ and $\omega_{1}$.
from standard photodetection theory,\cite{28} the detection probability
 is
proportional to the first-order correlation function of the field
\begin{equation}
\Gamma^{(1)}(t)=
\mbox{Tr}
\left\{ \hat{\rho}_i \hat{E}^{(-)}(x,t)\hat{E}^{(+)}(x,t) \right\},
\end{equation}
where
\begin{displaymath}
\hat{E}^{(+)}(x,t)=i\sum_{\omega_n}
\sqrt{\frac{\hbar\omega_n}{2V}}
a_{{\small\mbox{\boldmath $\epsilon_n$}},\omega_n}
\exp (-i\omega_n t+ik_n x),
\end{displaymath}
\vspace{-4mm}
\begin{displaymath}
\hat{E}^{(-)}(x,t)=-i\sum_{\omega_n}
\sqrt{\frac{\hbar\omega_n}{2V}}
a^{+}_{{\small\mbox{\boldmath $\epsilon_n$}},\omega_n}
\exp (i\omega_n t-ik_n x),
\end{displaymath}
and $V$ is the normalization volume.  At this stage it is more convenient
to use a formal normalization of the states in a finite volume (see
below).  We can even use unnormalized states.  With this definition the
probabilities of the measurement outcomes will also be unnormalized, but
since only the relative probability is important for the different
measurements, this lack of normalization is unimportant.
Measurements of the correlation function of the field (the instantaneous
intensity) $\Gamma^{(1)}(t)$ are a realization of the  $E_{0,1}$ and
$E(dt)$ measurements described above in the sense that the statistics of
the outcomes gives the same information on the states as the statistics of
the  $E_{0,1}$ and $E(dt)$ measurements.  A combination of measurements
using a fast photodetector and measurements using two narrow-band filters
and the same photodetector can provide information on the amplitude
$|f_{0,1}|$ and relative phase of the components $f_{0}$ and $f_{1}$,
which exhausts the information on the states (see also
Ref.~\cite{22}).

The probability $p$ of a photon being recorded in the time interval $(t,t+ dt)$
by an ideal photodetector is proportional to the field intensity 
$I(t)\propto\Gamma^{(1)}(t)$ (Ref.~\cite{28}):
\begin{equation}
p(t)dt\propto I(t)dt=\Gamma^{(1)}(t)dt.
\end{equation}
If the photodetector trigger time is $\tau_{\rm det}\ll
1/|\omega_1-\omega_0|$, this photodetector implements $E(dt)$ measurements
in the sense indicated above.  It can be seen from Eq.~(10) that for state
(9) the recording probability with allowance for Eqs.~(9)--(11) has the
form
\begin{eqnarray}
p(t)dt\propto I(t)dt&=& \Gamma^{(1)}(t)dt=
\left|
\sqrt{\omega_0}f_0
\exp \left[-i\omega_0(t-t_0)+\frac{ik_0L}{c}\right]
  \right.+\nonumber\\
&+&\left.
\sqrt{\omega_1}f_1  
\exp \left[-i\omega_1(t-t_0)+ \frac{ik_1L}{c}\right]\right|^2
\frac{dt}{2V},
\end{eqnarray}
where $k_{0,1}$ are the wave vectors corresponding to the frequencies
$\omega_{0,1}$ and $L$ is the length of the communication channel (we
assume that the measurement is made at a distance $L$ from the
transmitting end).

Measurements of the amplitude of the spectral components $f_{0,1}$ are
made using a pair of narrow-band filters which cut out the frequencies
$\omega_{0,1}$ prior to photodetection, and the same photodetector.  The
recording probability, in accordance with Eqs.~(9)--(11), does not depend on 
time:
\begin{eqnarray}
p_c(t)dt\propto \Gamma^{(1)}(t)dt&=&
\left\{
\begin{array}{lcr}
\displaystyle\frac{\hbar\omega_0}{V}|f_0|^2dt,
&E_0&\mbox{~measured}, \\[3mm]
\displaystyle\frac{ \hbar\omega_1}{ V}|f_1|^2dt,&E_1
&\mbox{~measured},\\
\end{array}
\right.\nonumber \\
p_{0,1}(t)dt\propto \Gamma^{(1)}(t)dt&=&
\left\{
\begin{array}{lcr}
\displaystyle\frac{ \hbar\omega_0}{ V}
\times 1\, dt,&E_0 &\mbox{measured for}\quad \hat{\rho}_{0},\\
[3mm]
0        &E_1  &\mbox{measured for}\quad \hat{\rho}_{0},\\
[3mm]
\displaystyle\frac{\hbar\omega_1}{ V}
\times 1\, dt,&E_1 &\mbox{measured for}\quad \hat{\rho}_{1},\\
[3mm]
0         &E_0 &\mbox{measured for}\quad \hat{\rho}_{1}.\\
\end{array}
\right. 
\end{eqnarray}

A qualitative version of a cryptosystem is shown in Fig.1.  Before
entering the line, the signal from a single-photon source is expanded into
a spectrum from which are cut either one of the frequencies ($\omega_0$ or
$\omega_1$), or both spectral components with the frequencies $\omega_0$
and $\omega_1$. The $E(dt)$ measurements are made using a fast
photodetector operating in a waiting mode.  In this mode the occurrence of
an event (its recording) will take place at a random time,  not chosen by
the experimentalist. This  differs from the $E(dt)$ measurements made at 
a 
time preselected by the experimentalist in the range $(t,t+dt)$;  the
probability of recording at this time is described by the density
$p_{c}(t)$.  In this case, the $E(dt)$ measurement cannot be understood as
a measurement by the fast photodetector which has an input diaphragm in
front of it which is opened in the interval $(t,t+dt)$.  This procedure
also corresponds to some measurement, but not to an $E(dt)$ measurement.
The integrated recording probability at time $T$ is given by
$$
P(T)=\int\limits^{T} dt\, p_{i}(t),\quad  i=c,0,1,
$$ 
from which the probability density in the time interval $(t,t+dt)$ can be
obtained by differentiating with respect to the upper limit.
The $E(dt)$ measurement essentially contains information on the
``interference'' of the different spectral component within a single
quantum state (information on the relative phase of the components with
the frequencies $\omega_{0}$ and $\omega_{1}$).  Thus, it is fundamentally
important that in different messages the state $\hat{\rho}_c$ is prepared
so that the relative phase of the spectral components is the same.
Otherwise, the time-oscillating (interference) component with the
frequency $\omega_1-\omega_0$ in the probability $p_{c}(t)$ will not be
reproduced at the same times in different attempts.  The problem of
preparing a single-photon state where the relative phase of the components
is fixed can be solved as follows.  Let us assume that a two-level system
exists with a spin-nondegenerate electron spectrum (for example, a quantum
dot with Coulomb interaction, see details given in Ref.~\cite{29}).

Resonant illumination with a square-wave pulse can transfer the system to 
the excited (quasi-steady) state.  We shall assume that the duration of
the square-wave pulse is much shorter than the radiative recombination
time ($\tau_{\pi}\ll \tau_R$).  The duration of the square-wave pulse can
be made substantially shorter than $\tau_{R}$  with some margin for not
impairing the condition of resonant illumination.\cite{29}  This implies
that the time of excitation $t_{0}$ is determined to within
$\sim\tau_{\pi}\ll\tau_R$.  After  the square-wave pulse has been switched
off, the free evolution of the system comprising an electron in the
excited state plus the electromagnetic field in the vacuum state leads to
recombination of the electron and the appearance of a single-photon packet
with the characteristic spectral width $\Delta\omega\approx 1/\tau_R$. The
single-photon packet is defined as\cite{30,31}
\begin{equation}
|1_f\rangle=
\sum_{k}f_ka^{+}_{\omega_k}|0\rangle,\quad \sum_{k}|f_k|^2=1.
\end{equation}
The average number of photons in the packet is
\begin{equation}
n=\langle 1_f|a^{+}_{\omega_k}a_{\omega_k}|1_f\rangle=1,
\end{equation}
and physically this implies that recording by an ideal wide-band
photodetector (which captures all the spectral components) leads to
triggering  with a probability of unity.  Recording by an ideal
narrow-band detector at the frequency $\omega_{n}$ leads to a probability
$|f_n|^2<1$ of triggering.

Since each system at time $t_{0}$ starts from the same state, in different
messages the single-photon packets are the same (the phase of all the
spectral components determined by the factors $\exp (-i\omega_i t_0)$ is
the same in different messages).  Cutting out two narrow spectral
components from the spectrum conserves their relative phase.  In fact, the
cutting of the spectral components is formally described as the action of
the projector\footnote{Strictly speaking, an ideal filter corresponds to 
a projector on the subspace of states with frequencies $\omega_0$,
$\omega_1$ and all the occupation numbers 
$E_{\omega_0}+E_{\omega_1}=\sum\limits_{n=0}^{\infty}(|n_{\omega_0}
\rangle\langle n_{\omega_0}|+|n_{\omega_1} \rangle\langle n_{\omega_1}| 
\rangle)$, but this does not alter the results. }
$$
E_0+E_1=(|1_{\omega_0}\rangle\langle 1_{\omega_0}|+
|1_{\omega_1}\rangle\langle 1_{\omega_1}|),
$$
after which the density matrix of the single-photon wave packet is
transferred to a new state (see Refs.~\cite{25,26}, and
\cite{32})
\begin{eqnarray}
\hat{\rho}_{in}(t)&=&
\left\{
\sum_{k}
\exp \left[-i\omega_k(t-t_0)\right]f_k|1_{\omega_k}\rangle\right\}\times
                \nonumber\\
&\times&
\left\{\sum_{k'}\langle 1_{\omega_k'}|f^{*}_{k'}
\exp \left[i\omega_k'(t-t_0)\right]\right\}
\rightarrow
\frac{1}{
\mbox{Tr}\left\{\hat{\rho}_{in}(t)(E_0+E_1)\right\}
}\,
(E_0+E_1)     \times
                \nonumber\\
&\times&
\left\{\sum_{k}
\exp \left[-i\omega_k(t-t_0)\right]f_k|1_{\omega_k}\rangle\right\}
\left\{\sum_{k'}\langle 1_{\omega_k'}|f^{*}_{k'}
\exp \left[i\omega_k'(t-t_0)\right]\right\}
(E_0+E_1)
 \rightarrow 
            \nonumber\\
& \rightarrow &
\frac{1}{
 |f_0|^2+|f_1|^2}\,
\left\{
\exp \left[-i\omega_0(t-t_0)\right] f_0 |1_{\omega_0}\rangle  +
\exp \left[-i\omega_1(t-t_0)\right]f_1|1_{\omega_1}\rangle
\right\}\times
                \nonumber\\
&\times&
\left\{
\langle 1_{\omega_0}|f^{*}_{0}\exp \left[i\omega_0(t-t_0)\right]+
\langle 1_{\omega_1}|f^{*}_{1}\exp \left[i\omega_1(t-t_0)\right]
\right\}.
\end{eqnarray}
Physically this implies that if an ideal wide-band photodetector is placed
after the filters, in a large number of repeated tests it will only be
actuated in the fraction
$\mbox{Tr}\left\{\hat{\rho}_{in}(t)(E_0+E_1)\right\}$ of the total number
of cases. 

The relative phase of the components with frequencies $\omega_{0}$ and
$\omega_{1}$ is determined by their phase at the preparation time which is
attainable in principle, as described above.  Thus, provided that
$\tau_{\pi}\ll\tau_R\ll1/|\omega_1-\omega_0|$ we can assume that in
different messages the temporal interference pattern stays in place.  In
different messages the interference pattern only ``floats'' to the extent
of the inaccuracy of the initial preparation time  $\delta t_0$ by the
amount $\delta t_0\le\tau_{\pi}\ll T=2\pi/|\omega_1-\omega_0|$ which is
substantially less than the period of the temporal interference pattern.

Assuming that the square-wave pulse duration is $\tau_{\pi}\sim
10^{-12}$\,s (see Ref.~\cite{29}) and the radiative recombination
time $\tau_R\sim 10^{-10}$\,s (in this case the spectral width of the
initial single-photon state is $\Delta\omega\sim 10^{10}$\,Hz), and
spectral components of width  $\sigma\approx 10^{7}$\,Hz separated by the
distance  $\delta\omega=|\omega_1-\omega_0|\sim 10^8$\,Hz are cut out
(which is still very far from the limits now attainable), the required
photodetector response speed is satisfied for 
$\tau_{\rm det}\sim10^{-9}$\,s.
The chain of inequalities $\tau_{\pi}\ll\tau_R\ll\tau_{\rm det}\ll 
1/\delta\omega$ is then satisfied with some margin.  The efficiency as a
result of cutting out narrow spectral components of width
$\sigma\sim10^7$\,Hz from a spectrum of width
$\Delta\omega\sim10^{10}$\,Hz is $\sim \sigma/\Delta\omega\sim 10^{-3}$.

However, at least the problem of a strictly single-photon source can be
solved in principle.
We shall estimate the accuracy in fixing the length of the communication
channel. Changes in the length of the  fiber-optic line also lead to
blurring of the interference pattern as a result of the presence of terms
with $k_{0,1}L$ in the exponent in formula (12).  The changes in the
relative phase of the spectral components as a result of variation of the
line length $\delta L$ should satisfy the condition
$$
|k_1-k_0|\delta L\approx|\omega_1-\omega_0|\delta L/c \ll 2\pi,
$$
and permissible variations of the line length should lead to a relative
phase shift much less than $2\pi$.  This gives the estimate
$$
\delta L\ll 2\pi c/\delta\omega\approx 10^2~{\rm cm},
$$ 
which is a fairly soft condition.

The interference pattern may also become blurred as a result of the
polarization vector rotating at different speeds in the different
frequency components.  However, if the position of the cable is fixed, the
interference pattern may be precalibrated.  In this case, any changes will
only be attributed to the different optical paths of the frequency
components, i.e., variation of the line length. This condition is clearly
noncritical.  The frequency dispersion of the dielectric constant of the 
optical fiber can also lead to smoothing of the amplitude of the
oscillations of the interference pattern.  The longer the line, the
stronger this smoothing.  However, estimates\cite{21} show that if the
width of the spectral components is $\sigma\approx 10^{7}$\,Hz, the
dispersion has an influence at far greater lengths than the attenuation.

Attenuation does not influence the secrecy of the system and only reduces
its efficiency by increasing the fraction of idle measurements.

The states with infinitely narrow spectral components analyzed above are
an idealization and are unsuitable for transmission along a communication
channel because of their formally infinite duration.  In real experiments
we can only prepare states with a finite line width (the preparation of
strictly monochromatic photon states would require a formally infinite
time).  The information states can be single-photon states of the type
(14) with Gaussian spectral densities
\begin{equation}
|1_{\omega_{0,1},c}\rangle=
\int\limits_{0}^{\infty}f_{0,1,c}(\omega)a^{+}(\omega)|0\rangle, 
\quad
\left[a(\omega),a^{+}(\omega')\right]=\delta(\omega-\omega')\hat{I},
\end{equation}
\vspace{-4mm}
\begin{equation}
E^{(+)}(x,t)=
\frac{1}{ \sqrt{2\pi} }
\int\limits_{0}^{\infty}
\exp \left [-i\omega \left(t- \frac{x}{c}\right)\right]
a(\omega)d\omega,
\end{equation}
\vspace{-4mm}
\begin{equation}
f_{0,1}(\omega)=
\frac{ 1}{ (2\pi\sigma^2)^{1/4} }
\exp 
\left[ -\frac{(\omega-\omega_{0,1})^2}{ 4\sigma^2} 
\right]
\exp (-i\omega t_0)
\end{equation}
and a control state containing both narrow-band Gaussian components with
amplitudes $f_{0}$ and $f_{1}$:
\begin{equation}
f_{c}(\omega)=
\frac{{\rm  const}}{ (2\pi\sigma^2)^{1/4}}
\left\{
f_0 
\exp \left[ -\frac{(\omega-\omega_{0})^2}{4\sigma^2} 
\right]+
f_1 \exp \left[ -\frac{ (\omega-\omega_{1})^2}{
4\sigma^2} \right]
\right\}\exp (-i\omega t_0),
\end{equation}
where the normalization constant is
\begin{equation}
{\rm const}=\left\{
|f_0|^2+|f_1|^2+
\sqrt{2}\mbox{Re}\left[ f_0f_1^*\right]
\exp \left[-\frac{(\omega_0^2+\omega_1^2-\omega_0\omega_1) }
{2\sigma^2} \right]
\right\}^{-1}.
\end{equation}
Measurements of narrow spectral components using suitable Gaussian filters
yield a weak time dependence of the measurement outcomes unlike the
previous analysis of strictly monochromatic states where the probability
of the outcome did not depend on time.  The corresponding probability
density of the results has the form
\begin{equation}
p(t)dt\propto I(t)dt=
2\sqrt{2\pi\sigma^2} \,
\exp \left[-2\sigma^2 (t-t_0-L/c)^2\right]dt,
\end{equation}
from which it follows in particular that the probability of recording by 
a
photodetector in the waiting mode only tends to unity if the waiting time
$T$ exceeds the reciprocal width of the spectrum ($T\geq 1/\sigma$).  This
factor is consistent with intuitive ideas on the prolonged recording time
of a narrow-band state. 

The probability density of the measurement outcomes for the control state
has the form
\begin{eqnarray}
&&
p_c(t)dt\propto I(t)dt=
{\rm const}\cdot 2\sqrt{2\pi\sigma^2}
\exp \left[-2\sigma^2 (t-t_0-L/c)^2\right]
\times
            \nonumber\\
&&\quad \times
\left\{
|f_0|^2+|f_1|^2+2\mbox{Re}
\left(
f_0f^*_1\exp \left[-i(\omega_0-\omega_1)(t-t_0-L/c)\right]
\right)
\right\}dt.
\end{eqnarray}
The interference oscillating component is well-defined under the condition
$\sigma\ll |\omega_1-\omega_0|$.

\section*{4.~~Conclusions}

Although various versions of quantum cryptosystems have been proposed, in
the author's view there is some indeterminacy associated with the
following.  The evidence of secrecy in quantum cryptography using two
nonorthogonal states proposed in Ref.~\cite{3} implies that the
states are stationary and belong to the same energy.  Otherwise, for
nonstationary states the projectors  $\overline{E}_0$ and $\overline{E}_1$
would differ at different times.  The nonorthogonality of the stationary
states implies that they correspond to the same energy.  Otherwise the
stationary states belonging to different energies would automatically be
orthogonal.  In this sense, the protocol for the stationary states exists
as it were outside time.  Attempts to introduce the time explicitly in the
exchange protocol\cite{14,17} still use reasoning from
Ref.~\cite{3} for stationary states as evidence of secrecy (see, for
example, Ref.~\cite{17}).  The stationary state are infinite in
time.  A similar situation arises here.  Evidence of secrecy for states
with infinitely narrow spectral densities, which are thus infinitely
extended in time, is also based on the reasoning given in
Ref.~\cite{3}.  For the case where the state space of the system is
infinite-dimensional (described by a continuous variable), evidence of
secrecy of the same degree of rigorousness as in Ref.~\cite{3} is
not obtained, as far as we are aware.  In this sense, the author takes the
view that evidence of secrecy is not obtained for real-time quantum
cryptosystems.

In conclusion, the author is grateful to B. A. Volkov, S. S. Nazin, S. T.
Pavlov, and I. I. Tartakovski\u{\i} for fruitful discussions.  This work
was supported by the Russian Fund for Basic Research (Project No.
96-02-19396).

\newpage
\begin{figure}[tb]
\caption{Qualitative version of cryptosystem.  The signal from a
single-photon source is directed to a ``prism'' beyond which is a screen
which transmits the signal either with the frequency $\omega_{0}$ (logic
zero, upper diaphragm open) or $\omega_{1}$ (logic one, lower diaphragm
open), or both frequencies (control signal, both diaphragms open).  At the
receiving end the measurement procedure is arranged similarly.  Upper
diaphragm open --- measurement of $E_{0}$, lower open --- measurement of
$E_{1}$, and both open --- measurement of $E(dt)$.  The diaphragms are
open or shut during an entire specific message. }
\end{figure}
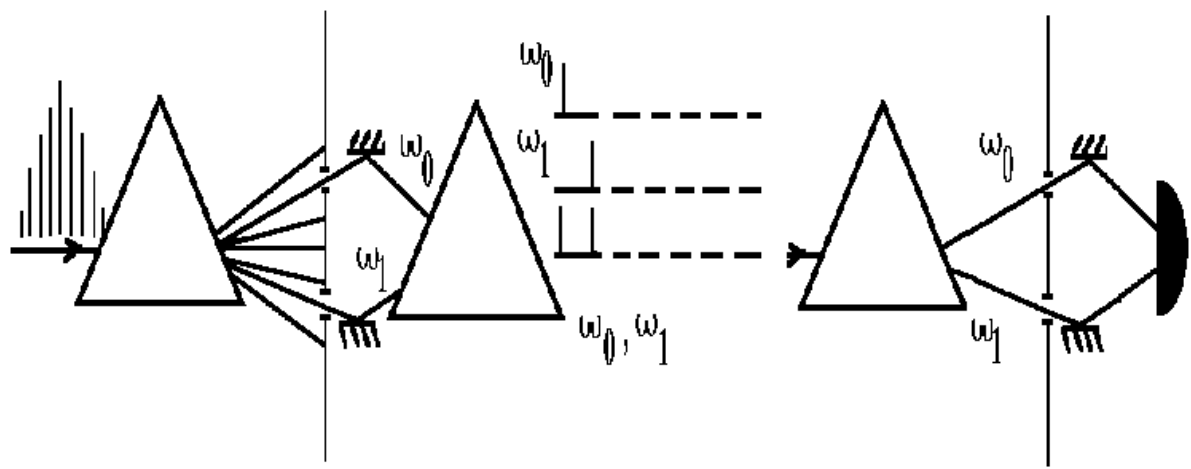
\end{document}